# Exploiting k-space - frequency duality in Fourier optics towards real-time compression-less terahertz imaging

Hichem Guerboukha, Kathirvel Nallappan, and Maksim Skorobogatiy*
École Polytechnique de Montréal
Montréal, Québec, H3T 1J4, Canada
*Corresponding author : maksim.skorobogatiy@polymtl.ca
2017-08-24

**Abstract:** We present theoretical formulation and experimental demonstration of a novel technique for the fast compression-less terahertz imaging based on the broadband Fourier optics. The technique exploits k-vector – frequency duality in Fourier optics which allows using a single-pixel detector to perform angular scan along a circular path, while the broadband spectrum is used to scan along the radial dimension in Fourier domain. The proposed compression-less image reconstruction technique (hybrid inverse transform) requires only a small number of measurements that scales linearly with the image linear size, thus promising real-time acquisition of high-resolution THz images. Additionally, our imaging technique handles equally well and on the equal theoretical footing the amplitude contrast and the phase contrast images, which makes this technique useful for many practical applications. A detailed analysis of the novel technique advantages and limitations is presented, as well as its place among other existing THz imaging techniques is clearly identified.

## Introduction

Terahertz frequency range (0.1 to 10 THz, wavelengths of 3 mm to 30 µm) has received a considerable attention over the years thanks to the prospect of numerous advanced imaging applications benefiting from the fact that many materials are semi-transparent to THz waves (for example polymers, plastic packaging, paper etc.) [1, 2]. Moreover, unlike X-rays, THz radiation is non-ionizing, thus, posing no risks to living beings. Many applications have been demonstrated in various applied fields such as security [3], biomedical [4], pharmaceutical [5], food industry [6], art conservation [7], etc. Despite all the interest and potential, many challenges remain that impede a widespread use of THz imaging.

One of the main limiting factor is the acquisition time of a THz image. Currently, spectral imaging is done using THz time-domain spectroscopy (THz-TDS) systems. The emitted THz pulse contains multiple frequencies and the detection is based on a time-domain sampling of the electric field, which provides direct access, through frequency Fourier transform, to the amplitude and the phase of the picosecond pulse.

There are two main types of the broadband THz detectors: 1) THz photoconductive antennas (THz-PCA) [8] and 2) THz detection based on electro-optic sampling (EOS) in nonlinear crystals [9]. Both techniques are highly sensitive, the THz-PCA generally performs better at frequencies below 3 THz with a higher signal-to-noise ratio thanks to the use of lock-in amplifiers and single pixel detectors, while EOS has a better sensitivity at higher frequencies [9-11] but lacks the ability of using the lock-in amplification when used together with CCD arrays. For spectral imaging, both methods share many similar challenges.

The first challenge is a slow time-domain pulse sampling typically based on a mechanical optical delay line. A potential solution to this problem is the asynchronous optical sampling that allows to forgo mechanical delay line and features repetition rates of several kHz [12]. However, this method is expensive as it uses two synchronized femtosecond lasers. A cheaper solution can be the use of fast rotary optical delay lines. These generally come either be in a prism [13-16] or a mirror [17-21] configuration.

The second challenge is a single-pixel nature of the THZ-PCA detectors. Some attempts were made recently to integrate several antennas on the same semiconductor chip [22-25]. For example, in [23], the authors demonstrated a 1D linear array of 15 pixels that was able to reduce the acquisition time from 9 hours to 36 minutes. However, low THz signals were reported and were mainly attributed to focusing optics that are complicated to manufacture. Additionally, dense integration of multiple antennas on a chip is problematic due to crosstalk and interference between them. In the EOS technology, on the other hand, it is possible to replace the single-pixel photodiode by a Charge-coupled device (CCD) [26]. However, due to impossibility of integrating lock-in amplifiers into the imaging setup, the loss in signal is severe, thus requiring additional data postprocessing such as the dynamic subtraction technique [27] or the averaging over multiple frames.

Together, the two abovementioned challenges are currently major limiting factors that prevent the proliferation of THz spectral imaging to real-time imaging applications. With a single pixel detector, spectral imaging experiments are generally performed by physically moving a sample in the focal plane of a focusing optics. When performing time-domain sampling using an optical delay line (based on a linear micro-positioning stage), a single pixel with a spectral resolution of ~1-3 GHz is typically acquired in 1-5 seconds. Therefore, even a low spatial resolution 32x32 spectral image of 1024 pixels takes ~1h to acquire.

Due to complexity in building high sensitivity THz multipixel arrays, the general trend in the THz community is to reduce the number of pixels required to reconstruct an image, by using advanced signal processing techniques such as Compressive sensing (CS) [28]. In application to imaging, the CS theory is based on the assumption that most objects have a sparse representation in a given basis. This allows to efficiently reconstruct a $N \times N$ image with less than $N^2$ measurements. The very first demonstration of CS theory applied to THz imaging was done using Fourier optics [29]. With a moving single-pixel detector, the Fourier transform of an object placed in the focal plane of a lens was measured. Then, using a CS algorithm, an image of 4096 pixels was reconstructed with only 500 measurements (12%) at a fixed frequency of 0.2 THz. However, this approach required mechanical movement of the detector. In a second work [30], the same authors used a set of random binary metal masks at the position of the object and a fixed single-pixel detector. The binary masks form a basis using which a 1024 pixels' image was reconstructed using 300 (29%) to 600 (59%) measurements at a fixed frequency of 0.1 THz.

The possibility of reducing the number of measurements and the prospect of using only a single-pixel detector resulted in a spurt of activity in CS applied to THz imaging. Thus, instead of using random masks, in [31], the authors proposed a set of optimized masks. In an attempt for real-time practical application and to avoid the manual change of the masks, Shen *et al.* used a spinning-disk with holes as the mask [32]. More recently, researchers have used an optically controlled silicon mask in the THz path [33. When illuminated with visible light, the mask made of high-resistivity silicon becomes opaque to THz. Then, using a digital micro-mirror device, the authors created various patterns similar to the binary metal masks. A THz image of 1024 pixels was then obtained with only 63 measurements in 2 seconds, with the main limitation being the speed of the digital micro-mirror used in the photomask generation. In [34], the same group used an electronically-controlled spatial light modulator in the THz beam. They demonstrated image acquisition at 1 Hz by using 45 masks/second. They also recorded a video of a metallic moving object at a speed of 1.8 mm/s. Furthermore, most of the previous research in CS was done on reconstructing amplitude modulated images predominantly in the form of binary metal masks with cutouts. However, the practical applications frequently require phase sensing. For example, when thinking about quality control, a scratch on the surface of a plastic sheet will result in a weak amplitude contrast, but a strong phase contrast.

In this work, we present a novel compression-less imaging technique based on the k-space/frequency duality in Fourier optics, that is capable of reconstructing amplitude and phase contrast THz images using only a small number of measurements proportional to the linear size of the object rather than its area. As noted in [35], in Fourier space a spectral frequency can be equated with a spatial frequency, which opens the possibility of substituting sampling over 2D k-space by sampling over 1D k-space and frequency. Therefore, we first use a single pixel THz-PCA detector and a lock-in amplifier to record the broadband time pulses at some strategically chosen points in the Fourier space. Then, we use a novel hybrid inverse transform developed in our group to reconstruct both amplitude, and (for the first time) phase images from the time traces collected in the Fourier space. A detailed analysis of the amplitude and phase image resolution is then presented using experimental measurements and numerical data. Unlike the techniques based on Compressive sensing theory, our method is lossless (in term of information) as both spatial and spectral data is used for image reconstruction. Based on solid mathematical formulation, our hybrid inverse transform can be equally well applied both to amplitude and phase imaging.

## Results

**Hybrid image reconstruction algorithm using a generalized Fourier optics approach**

In what follows, we first introduce a hybrid image reconstruction algorithm that is based on a generalization of Fourier optics approach to broadband pulses. The Fourier optics theory states that, for a fixed frequency $\nu$, a field profile $S(x, y, \nu)$ generated by an object that is placed in the front focal plane (object plane) of a convex lens is Fourier-transformed according to [36]:

$$U(\xi, \eta, \nu) = \frac{\nu}{jcF} \iint dxdy \cdot S(x, y, \nu) \exp\left[-\frac{j2\pi\nu}{cF}(x\xi + y\eta)\right] \quad (1)$$

where $F$ is the lens focal distance, and $c$ is the speed of light. The Fourier transform $U(\xi, \eta, \nu)$ of the original field can be measured directly in the lens's back focal plane (also known as the Fourier plane), where $(\xi, \eta)$ are the Cartesian coordinates of the observation point (see schematics in Fig. 1a). Moreover, the original field distribution $S(x, y, \nu)$ generated by the object can be reconstructed using the inverse Fourier transform:

$$S(x, y, \nu) = \frac{j\nu}{cF} \iint d\xi d\eta \cdot U(\xi, \eta, \nu) \exp\left[+\frac{j2\pi\nu}{cF}(x\xi + y\eta)\right] \quad (2)$$

The spatial frequencies, also known as components of the k-space, are related to the $(\xi, \eta)$ coordinates as:

$$k_\xi = \frac{\xi\nu}{cF} \qquad k_\eta = \frac{\eta\nu}{cF} \quad (3)$$

When using broadband pulses, a raster scanning in the Fourier plane (Fig. 1b) results in acquisition of a hyperspectral cube where for each frequency there is an image in the k-space (Fig. 1c). However, as the k-components (3) are proportional to the frequency $\nu$, one quickly arrives at an intriguing idea of sampling the k-space using the broadband nature of the THz-TDS pulse, rather than by a mechanical scanning of the k-space by displacing the detector. Indeed, by fixing a detector at a fixed position with coordinates $(\xi_0, \eta_0)$ and using a broadband light $\nu \in (\nu_{min}, \nu_{max})$ one can sample a linear segment of the k-space described by the equation:

$$k_\eta = \frac{\eta_0}{\xi_0} k_\xi \qquad \text{with } k_\xi \in \frac{\xi_0}{cF} \cdot (\nu_{min}, \nu_{max}) \qquad (4)$$

By changing the ratio $\eta_0/\xi_0$, the whole k-space can be sampled. The simplest way to change this ratio is to measure several points along a circle of fixed radius $\rho_0$ in the $(\xi, \eta)$ plane (Fig. 1d and 1e). Mathematically, for further consideration, it is more convenient to write the Fourier transform (1) in polar coordinates:

$$U(\bar{\rho}, \nu) = \frac{\nu}{jcF} \iint d\phi r dr\, S(\bar{r}, \nu) \exp\left[-\frac{j2\pi\rho}{cF} \bar{r} \cdot \bar{\rho}\right] \qquad (5)$$

where $\bar{\rho}$ is a vector position in the observation point in the Fourier plane, while $\bar{r}$ is the vector position in the object plane. The inverse Fourier transform (2) is then written as:

$$S(\bar{r}, \nu) = \frac{j\nu}{cF} \iint d\theta \rho d\rho\, U(\bar{\rho}, \nu) \exp\left[+\frac{j2\pi\rho}{cF} \bar{r} \cdot \bar{\rho}\right]$$
$$= \iint d\theta \left(\frac{j\nu\rho}{cF}\right) d\left(\frac{j\nu\rho}{cF}\right) \frac{U(\bar{\rho}, \nu)}{\frac{j\nu}{cF}} \exp\left[+2\pi \left(\frac{j\nu\bar{\rho}}{cF}\right) \cdot \bar{r}\right] \qquad (6)$$

As it can be seen from the integral in equation (6), the frequency $\nu$ of the probing wave and the distance of the detector from the origin $\rho$ in the Fourier plane are entering the formulation in a symmetrical fashion. Therefore, in principle, the integral over the spatial coordinate in the Fourier plane $\bar{\rho}$ at a fixed frequency $\nu_0$ can be replaced by an integral over the frequency $\nu$ at a fixed radius $\rho_0$. This means that instead of fixing the frequency of the probing wave and recording the Fourier image by 2D scanning of a point detector, we can rather only scan along a single 1D circular path of radius $\rho_0$, while using the broadband spectrum of the probing pulse to sample the k-space along the radial direction. This way, the full 2D raster-scan at a fixed frequency over the Fourier plane is avoided and is substituted by a time domain scan along a single circle.

Mathematically, we, therefore, define a new hybrid inverse transform by substituting the integration over the 2D Fourier space in (6) by a hybrid integration over 1D spatial coordinate and frequency:

$$\tilde{S}(\bar{r}) = \iint d\theta\, \nu d\nu \left(\frac{j\rho_0^2}{cF\nu} U(\rho_0, \nu, \theta)\right) \exp\left[+\frac{j2\pi\nu}{cF} \bar{r} \cdot \overline{\rho_0}\right] \qquad (7)$$

It is important to note that the image $\tilde{S}(\bar{r})$ reconstructed using equation (7) is different from the original image $S(\bar{r}, \nu)$ as given by the standard inverse Fourier transform equation (6). Indeed, $S(\bar{r}, \nu)$ is a hyperspectral image which can be different for different frequencies $\nu$. In contrast, $\tilde{S}(\bar{r})$ is a compounded image that incorporates information from all the frequencies sampled by the pulse. Therefore, one must recognize that although the definition of the hybrid inverse transform (7) is derived using physical arguments, its final form should be considered as a mathematical abstraction, and therefore can be generalized even further as follows:

$$\tilde{S}(\bar{r}) = \iint d\theta\, \nu d\nu \left(\frac{j\rho_0^2}{cF\nu} \frac{U(\overline{\rho_0}, \nu)}{U_{ref}(\nu)}\right) \exp\left[+\frac{j2\pi\nu}{cF} \bar{r} \cdot \overline{\rho_0}\right] \qquad (8)$$

where $U_{\text{ref}}(\nu)$ is a certain frequency dependent reference function responsible for the normalization of the integral (8). As we will see in what follows, this normalization step is crucial for correct phase and amplitude retrieval, as well as correct interpretation of the image given by (8). In what follows, we demonstrate that in several important cases of amplitude and phase masks, properly normalized hybrid transform (8) recovers the original amplitude and phase information.

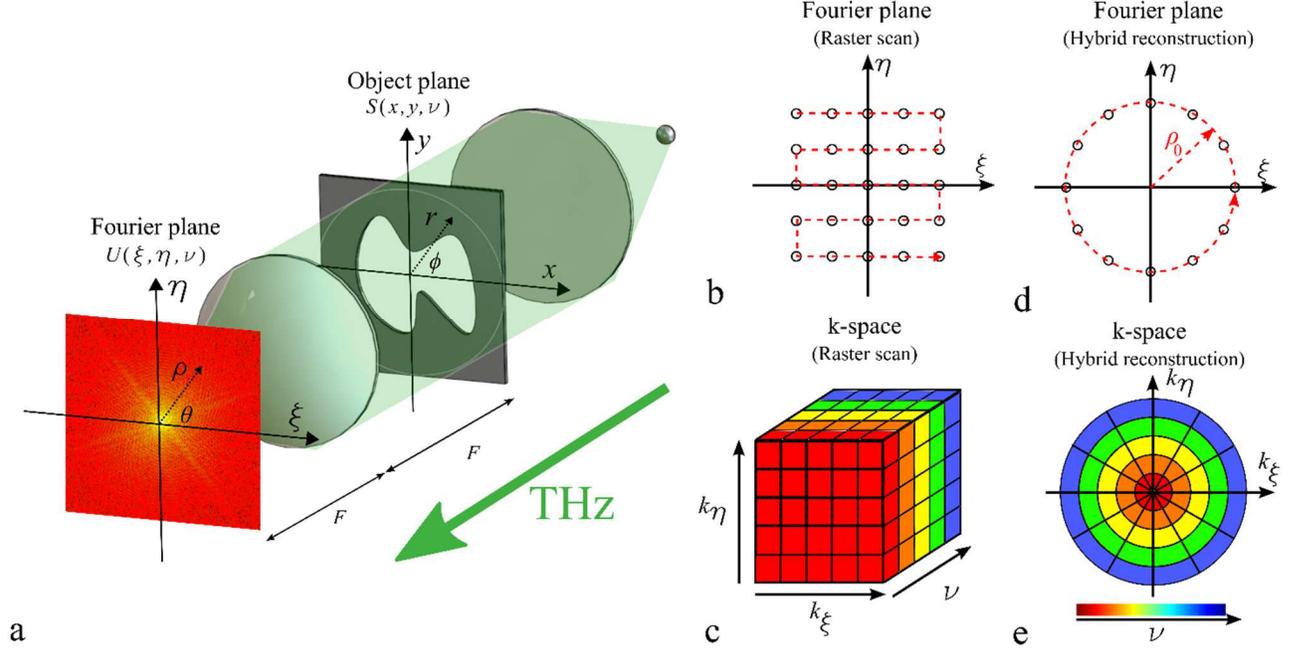

**Figure 1. Hybrid image reconstruction algorithm.** (a) Schematic of the object plane and the Fourier plane. (b) Raster scanning on a 2D Cartesian grid in the Fourier plane and (c) corresponding hyperspectral k-space cube. (d) Hybrid reconstruction algorithm with a 1D circular scan in the Fourier plane and (e) corresponding k-space inferred using spectral information.

**Amplitude masks**

First, we consider the case of an object in the form of an amplitude mask. The resultant image $S(\bar{r}, \nu)$ is assumed to be space-frequency separable, and in the object plane is defined as:

$$S(\bar{r}, \nu) = S(\bar{r})E(\nu) \qquad (9)$$

A typical example of this imaging modality would be a binary mask represented by an opaque screen with a cut-out pattern (image), which is illuminated with a pulsed light. In this case, $S(\bar{r})$ is a function that has only two values 1 or 0 depending whether it corresponds to the cutout or the opaque part of the screen. Here, $E(\nu)$ is the frequency dependent field of the probing pulse. Furthermore, as a reference function $U_{\text{ref}}(\nu)$ in the hybrid inverse transform (7) we take the normalized THz-TDS trace as measured by a point detector in the center of the Fourier plane ($\rho_0 = 0$) with the amplitude mask still present in the object plane:

$$U_{\text{ref}}(\nu) = (jcF/\nu) \cdot U(0, \nu). \qquad (10)$$

In practice, we first perform a time domain measurement in the origin of a Fourier plane, then $U(0, \nu)$ is found using a frequency Fourier transform of the measured time-domain pulse. Using equation (5), one can also write (10) as:

$$U_{\text{ref}}(\nu) = \frac{jcF}{\nu} U(0,\nu) = E(\nu) \iint d\bar{r}\, S(\bar{r}) \tag{11}$$

In this case, the hybrid inverse transform becomes:

$$\tilde{S}(r,\phi) = \iint d\theta d\nu \left[\nu \left(\frac{\rho_0}{cF}\right)^2 \frac{U(\rho,\theta,\nu)}{U_0(0,\nu)}\right] \exp\left[+\frac{j2\pi\rho_0}{cF} \bar{r}\cdot\overline{\rho_0}\right] \tag{12}$$

Note that the reference $U_{\text{ref}}(\nu)$ is chosen to contain an additional multiplicative factor $jcf/\nu$ that is necessary in order to ensure that the reconstructed image $\tilde{S}(r,\phi)$ is proportional to the original one $S(\bar{r},\nu)$. In fact, from the detailed mathematical analysis presented in the Methods section, it follows that in the case of amplitude masks, the hybrid inverse transform results in the original image normalized by the constant factor proportional to the area of the image:

$$\tilde{S}(r,\phi) = \frac{S(r,\phi)}{\iint S(r,\phi) d\phi r dr} \tag{13}$$

As an example, in Fig. 2, we experimentally demonstrate imaging of a metallic mask with a cutout in the form of a Canadian maple leaf (Fig. 2a). Our intent here is to demonstrate reconstruction of a complex image using the broadband spectrum and very few pixels in the Fourier plane. For comparison, we first perform regular imaging by recording the field distribution in the Fourier plane by raster scanning with a point detector on a 102x102 mm grid with 1.5 mm resolution (4624 pixels). At each pixel, a full frequency spectrum is recorded, thus a complete hyperspectral image is acquired. As an example, in Fig. 2b and 2c we present image amplitude and phase distributions in the Fourier plane at a particular frequency of 0.57 THz. Reconstruction of the original image is then performed using standard inverse Fourier transform in Cartesian coordinates (equation 2). The result contains both amplitude and phase information for every frequency. In Fig. 2d, we present amplitude distribution of a reconstructed image at 0.57 THz that corresponds to the amplitude and phase distributions in the Fourier plane presented in Fig. 2b and 2c.

We now show that the hybrid inverse transform presented in this section (equation 12) can reconstruct a simple binary image with considerably fewer pixels. Thus, in Fig. 2e and 2f, we present the recorded amplitude and phase distributions of the image in the k-space using only 180 pixels in the Fourier plane. These points are sampled on a circle of radius $\rho_0 = 25$ mm with the center at the origin. The traces are acquired at equal angular intervals. The data is then normalized by the complex spectrum recorded at the origin of the k-space (expression (11)). The k-space at Fig. 2e and 2f is constructed by interpreting the spectral data as components of the k-space with equation 3. As expected, this k-space compares remarkably with the k-space obtained through raster-scanning (Fig. 2b and 2c). Finally, the image is reconstructed numerically using integration in polar coordinates of equation 12 (Fig. 2g). Additionally, in Fig. 2h and 2i, we study the quality of the reconstructed image when using different number of pixels on the circle, between 45 and 20. We note that even when using as little as 20 pixels the maple leaf can still be clearly recognized.

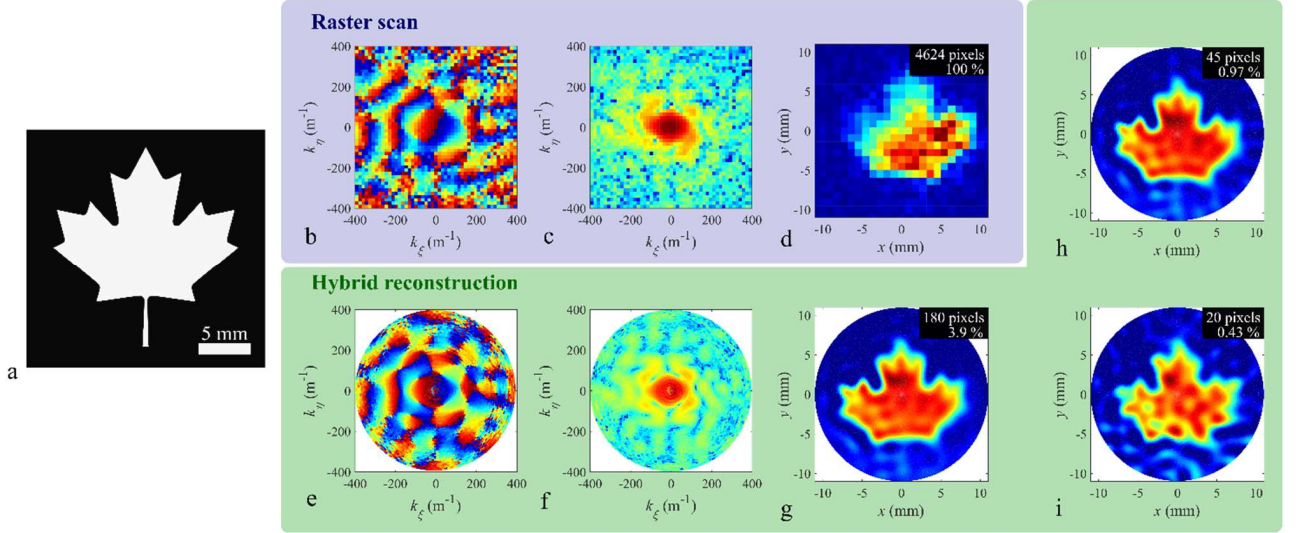

**Figure 2. Reconstruction of a binary amplitude image in the form of a maple leaf cutout in the metallic plate.** (a) Schematic of the maple leaf cutout. Standard raster scanning: (b) Amplitude and (c) phase of the k-space at a single frequency of 0.57 THz (4624 pixels). (d) Image reconstruction using standard inverse Fourier transform (2). Image reconstruction using hybrid inverse transform (12): (e) Inferred k-space amplitude distribution and (f) phase distribution using spectra of the THz time traces acquired at 180 pixels positioned along a circle of radius $\rho_0 = 25$ mm around the origin of Fourier plane. (g) Image reconstructed using hybrid inverse transform (12) with 180 pixels, (h) 45 pixels and (i) 20 pixels.

**Phase masks**

Second, we study the case of an object in the form of a dispersion-less phase mask with no absorption. As an example, we can consider imaging scratches on the surface of a transparent material plate/film surrounded by air (see Fig. 3a). In this case, the optical path of the probing light in the object plane is given by:

$$\Delta(\bar{r}) = \Delta_0 - \mu(\bar{r}) = [L_a n_a + L_m n_m] - [(n_m - n_a)h(\bar{r})] \qquad (14)$$

where $n_a$ and $n_m$ are the frequency-independent refractive indices of the air and the material, $L_a$ and $L_m$ are the distances travelled in the air and in the material, while $h(\bar{r})$ is the spatially dependent depth of the scratch. In this case, the field distribution in the object plane can be written as:

$$S(\bar{r}, \nu) = S(\bar{r})E(\nu) \exp[j2\pi\nu(\Delta_0 - \mu(\bar{r}))/c] \qquad (15)$$

where $S(\bar{r})$ defines the slow amplitude variation of the aperture-limited beam in the object plane, $E(\nu)$ is the frequency dependent field of the probing pulse, and $\mu(\bar{r}) = (n_m - n_a)h(\bar{r})$ is the optical path differential (a scratch, for example) across the object plane that we want to image.

In the case of phase masks, we define the reference $U_{\text{ref}}(\nu)$ in the hybrid inverse transform (8) using a THz pulse recorded at the origin of the k-space ($\rho_0 = 0$) and measured using a reference sample without the scratch ($h(\bar{r}) = 0$):

$$U_{\text{ref}}(\nu) = jcF\ U(0, \nu) = \nu\ E(\nu) \exp[j2\pi\nu\Delta_0/c] \iint d\bar{r}\ S(\bar{r}) \qquad (16)$$

Note that unlike in the case of the reference (11) used for amplitude masks, there is no division by the frequency in (16). This is an important difference with the case of amplitude masks that is detailed in the Methods section. The hybrid inverse transform (8) then becomes:

$$\tilde{S}(r,\phi) = \iint d\theta d\nu \left[\left(\frac{\rho_0}{cF}\right)^2 \frac{U(\rho,\theta,\nu)}{U_0(0,\nu)}\right] \exp\left[+\frac{j2\pi\rho_0}{cF}\bar{r}\cdot\overline{\rho_0}\right] \quad (17)$$

In the Methods section, we demonstrate that the imaginary part of the reconstructed image (17) is directly proportional to the optical path differential in the object plane:

$$\text{Im}\{\tilde{S}(\bar{r})\} = -\frac{2\pi}{c} \frac{S(\bar{r})}{\iint d\bar{r}\, S(\bar{r})} (n_m - n_a) h(\bar{r}) \quad (18)$$

where the term $S(\bar{r})/\iint d\bar{r}\, S(\bar{r})$ represents the slow-varying amplitude of the incident aperture-limited beam that can be measured independently (if desired) using the binary mask method described earlier. Moreover, we find experimentally, that our imaging approach given by the hybrid inverse transform (17) is so sensitive that it can readily detect phase nonuniformities in the phase distribution across the imaging plate. Such nonuniformities can be due to defects in the substrate geometry and composition, substrate misalignment, or even due to phase variation in the wavefront of the probing beam. Mathematically this means that instead of a constant phase $\Delta_0$ used in (15) we have to assume a somewhat nonuniform phase distribution $\Delta_0(\bar{r})$ across the surface of a reference substrate. Therefore, to improve the quality of phase imaging, we find it beneficial to substract the phase image of a pure substrate from the phase image of a phase mask written on a similar substrate. The phase image of a substrate $\text{Im}\{\tilde{S}_0(\bar{r})\}$ is reconstructed in exactly the same fashion as a phase image $\text{Im}\{\tilde{S}(\bar{r})\}$ of a phase mask written on a similar substrate by performing exactly the same steps and using the same reference (in both cases) as described above.

As an example, we demonstrate imaging of the Greek letter $\pi$ inscribed as a 100-μm deep engraving in a $L_m = 1$ mm thick slab of photosensitive resin (*PlasCLEAR*) printed using a stereolithography 3D printer (*Asiga® Freeform PRO2*) (Fig. 3a and 3b). This polymer is transparent in the THz region and has an almost constant refractive index of $n_m = 1.654$ [37]. The amplitude and the phase distributions of the image in the k-space are presented in Fig. 3c and 3d. As in the case of amplitude masks, these were inferred using the k-space/frequency duality in Fourier optics. In what follows, our goal is to map the height $h(\bar{r})$ of the engraving. First, we compute the imaginary part of the hybrid inverse transform $\text{Im}\{\tilde{S}(\bar{r})\}$ (18), which is shown in Fig. 3d. There, the $\pi$ engraving is clearly visible but it is surrounded by a halo due to somewhat nonuniform substrate used in the experiment. Two additional steps are required to extract the absolute value of the scratch depth $h(\bar{r})$. First, as described above we retrieve the phase image of a somewhat nonuniform substrate without the scratch $\text{Im}\{\tilde{S}_0(\bar{r})\}$, which is shown in Fig. 3e. Then, the phase variation across the reference substrate is removed from the image with the scratch by subtracting the two phase images $\text{Im}\{\tilde{S}(\bar{r})\} - \text{Im}\{\tilde{S}_0(\bar{r})\}$, the result is shown in Fig. 3f. There, we can see that the phase image has been considerably improved and only the letter $\pi$ is visible. To extract the absolute value of the scratch depth, we need to further divide the phase image $(\text{Im}\{\tilde{S}(\bar{r})\} - \text{Im}\{\tilde{S}_0(\bar{r})\})$ by the slow varying amplitude of the aperture-limited beam $S(\bar{r})/\iint d\bar{r}\, S(\bar{r})$. As mentioned earlier, this can be found by performing hybrid inverse transform (equation 12) while treating the substrate without a scratch as an amplitude mask. Finally, the height distribution can be extracted using equation (18) and as seen from the Fig. 3h, the $\pi$ symbol is an engraving of depth around 100 μm.

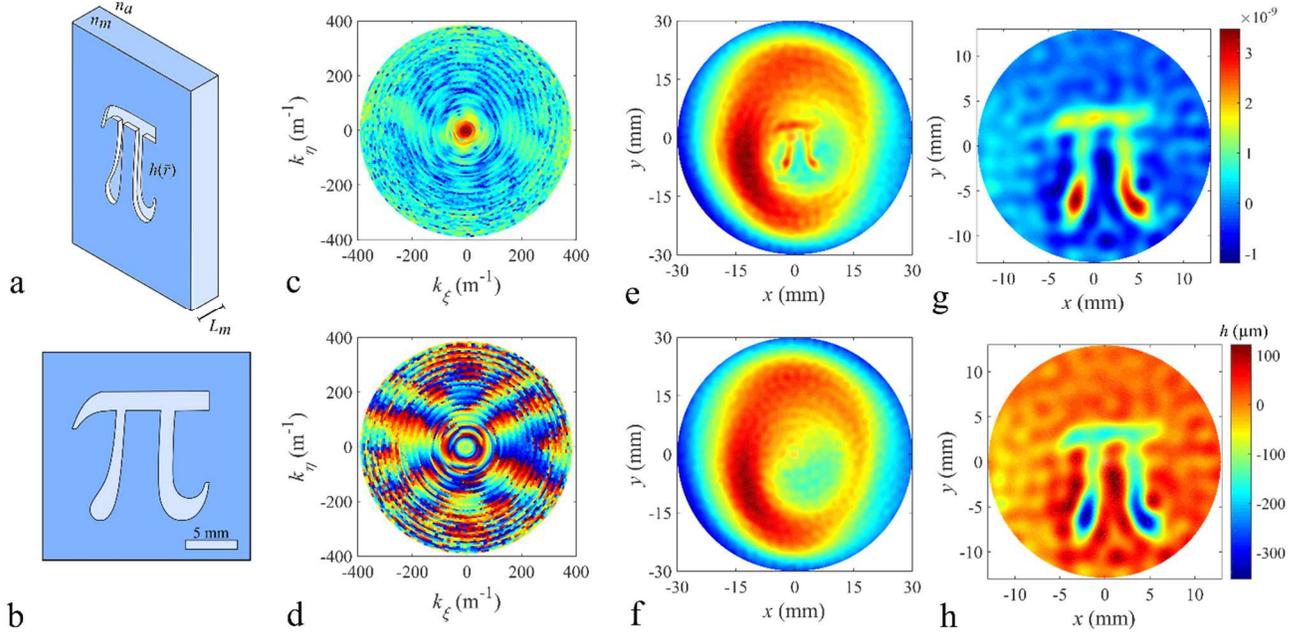

**Figure 3. Reconstruction of a phase contrast image in the form of the shallow engraving of the Greek letter $\pi$ onto a slab of transparent plastic.** (a),(b) Schematic of the sample $n_a = 1$, $n_m = 1.654$ and $L_m = 1$ mm. Image reconstruction using hybrid inverse transform (17): (c) Inferred k-space amplitude distribution and (d) phase distribution using spectra of the THz time traces acquired at 180 pixels positioned along a circle of radius $\rho_0 = 25$ mm around the origin of Fourier plane. (e) Reconstructed phase image of a substrate with engraving $\text{Im}\{\tilde{S}(\bar{r})\}$ and (f) without the engraving $\text{Im}\{\tilde{S}_0(\bar{r})\}$. (g) Improved phase image of the engraving $\text{Im}\{\tilde{S}(\bar{r})\} - \text{Im}\{\tilde{S}_0(\bar{r})\}$. (h) Reconstructed depth of the engraving from equation 18 and a supplementary beam amplitude measurement.

## Discussion

### Image resolution

First, we quote the resolution of a standard Fourier optics technique when using raster scanning in the Fourier plane. In this case the minimal achievable resolution in the object plane follows the Fourier transform properties. Particularly, using the Nyquist theorem, the minimal resolution achievable is $\delta r = 0.5/k_{\max}$ where $k_{\max}$ is the maximal spatial frequency sampled in the k-space. When using raster scanning, the resolution is therefore limited by the grid size $\Delta \xi$ in the Fourier plane. For example, in the $x$ direction, with the definitions in Fig. 4a, 4b, the resolution is $dx = 0.25/\Delta k_\xi = 0.25\ \lambda F/\Delta \xi$, where $\Delta \xi$ is the grid size in the Fourier plane. Therefore, the resolution improves when using smaller wavelengths and larger grid size. In the meantime, the grid spacing in the k-space $dk_x$ limits the field of view (maximal image size) such as $\Delta x = 1/dk_x = \lambda F/d\xi$, where $d\xi$ is the grid spacing in the Fourier plane.

In the hybrid inverse transform, $k_{\max}$ is a function of the maximal radial position of the detector $\rho_{\max}$ and the maximal THz frequency $\nu_{\max}$ of the pulse, so that $k_{\max} = \nu_{\max}\rho_{\max}/cF$. Therefore, the minimal achievable resolution is set by the Nyquist theorem to $\delta r = 0.5cF/\nu_{\max}\rho_{\max} = 0.5\ \lambda_{\min}F/\rho_{\max}$, where $\lambda_{\min} = c/\nu_{\max}$.

In the case of amplitude masks, as shown in the Supplementary section 4, the resolution of the hybrid inverse transform algorithm (equation 12) follows closely the Nyquist theorem limit $\delta r = 0.5\ \lambda_{\min}F/\rho_0$, where $\rho_0$ is the radius of the circle along which the time-domain data is sampled. Particularly, we find that the hybrid inverse transform acts as a linear smoothing filter that simply averages the image inside a circle of radius $\sim \delta r$ defined above.

In the case of phase masks, as shown in the Supplementary section 5, the resolution of the hybrid inverse transform algorithm (equation 17) is somewhat more complex. First, the resolution in the object plane is limited by the Nyquist theorem. Second, a spatially dependent correction term proportional to the local optical path variation due to surface inhomogeneity (presence of the engraving) must be added to the resolution: $\delta r(\bar{r}) = [0.5\, \lambda_{min} + h(\bar{r})(n_m - n_a)] \cdot F/\rho_0$, where $h(\bar{r})$ is the local height of the scratch, while $n_m$ and $n_a$ are the refractive indices of the substrate material and surrounding air. In order to experimentally demonstrate the effect of both the radial position of the detector and the THz bandwidth on the image resolution, we present in Fig. 4 imaging of the paper cutout in the form of a snowflake fabricated using a commercial paper puncher (Fig. 4c). The paper cutout can be considered as a phase mask due to subwavelength thickness of the paper and its low absorption, therefore the hybrid inverse transform in the form of equation (17) is used. In Figures 4d to 4i, we summarize our measurements by presenting the reconstructed phase images as a function of increasing THz frequencies ($\nu_{max}$) and radial positions ($\rho_0$). From these images, we conclude that the resolution of the system can be enhanced either by using spectrally broader THz pulses or by positioning the detector as far from the origin of the Fourier plane as possible. Indeed, by increasing the THz bandwidth, the resolution of the image clearly improves as it can be seen qualitatively by comparing Fig. 4d, 4e, 4f and Fig. 4g, 4h, 4i. Moreover, by positioning the point detector at a larger radial position $\rho_0$, the image resolution again is enhanced as seen by comparing Fig. 4d to Fig. 4g, Fig. 4e to Fig. 4h and Fig. 4f to Fig. 4i. There, the small dendrites in the snowflake become resolvable when the radial position $\rho_0$ is increased. We note that improvement in the image resolution stops when increasing the radial position $\rho_0$ of the detector beyond the size of the limiting aperture of the optical system (lens size).

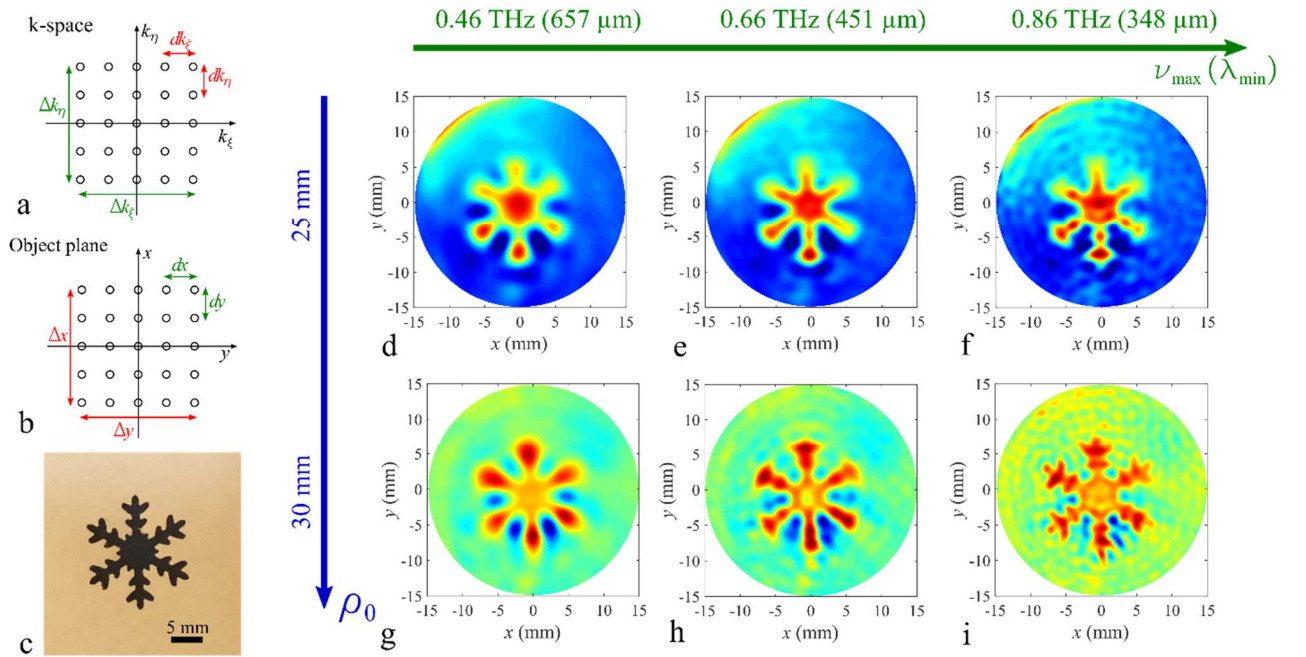

**Figure 4. Impact of the THz bandwidth and the radial position of the detector on phase image resolution.** Schematics of the (a) k-space and (b) the object plane showing relations between resolutions and image sizes in both planes. (c) Photograph of the phase mask in the form of a snowflake cutout in very thin paper (~100 μm) paper. Reconstructed phase images using equation (17) with $\rho_0 = 25$ mm and (d) $\lambda_{min} = 657$ μm ($\nu_{max} = 0.46$ THz), (e) $\lambda_{min} = 451$ μm ($\nu_{max} = $

0.66 THz), and (f) $\lambda_{min} = 348$ μm ($\nu_{max} = 0.86$ THz). Reconstructed images $\text{Im}\{\tilde{S}(\bar{r})\}$ using equation (17) with $\rho_0 = 30$ mm and (g) $\lambda_{min} = 657$ μm ($\nu_{max} = 0.46$ THz), (h) $\lambda_{min} = 451$ μm ($\nu_{max} = 0.66$ THz), and (i) $\lambda_{min} = 348$ μm ($\nu_{max} = 0.86$ THz).

**Advantages and limitations of the hybrid imaging algorithm**

Now, we would like to comment on the prospects of applying the presented imaging method in the context of industrial-strength real-time THz imaging systems. The main advantage of our method is its ability to reduce significantly the image acquisition time, while retaining the high signal-to-noise ratio offered by the TDS-THz setups based on photoconductive antennas, albeit with the loss of spectral data. In fact, the acquisition time of our method scales proportionally to the object linear size rather than the object area, which is the case in standard 2D raster scanning setups.

Moreover, as our system is based on the THz-TDS setup, the same imaging system can perform both hybrid imaging of the whole object (with the loss of the spectral data), followed by a more precise hyperspectral imaging using slow raster scan of a smaller target area of an object. Both amplitude imaging modality and phase imaging modality are supported by the hybrid imaging setup, thus allowing efficient imaging of the objects with high intensity contrast (amplitude modality) encountered for example in objects with deep carvings or cutouts, as well as low intensity contrast (phase modality) encountered for example in objects with small scratches or imperfections on their surface.

Additionally, we note that hybrid imaging based on the k-space/frequency duality opens a realistic approach to completely forgo mechanical scanning in high speed 2D THz imaging systems. Indeed, we note that the main reason that the standard 2D raster scanning is so slow is in the large part due to the mechanical scanning of the sample. Thus, with the current state of the art in micro-positioning stages, a reliable line sampling (back and forth along a single line of a ~5 cm long image) can be done with ~1 Hz rates. Therefore, even a modestly large image containing 100 lines would take several minutes to acquire. At this point we are not even talking about the hyperspectral imaging, but rather about imaging with a monochromatic THz source. Within our approach, we require acquisition of the full THz time traces along a single circle in Fourier space. Currently, this is accomplished using a 1D scan with a point detector. Alternatively, mechanical scanning of a point detector along a circle in the k-space can be forgone by adopting a 4f system layout (instead of 2f system used in our work), while using absorbing photomasks in the Fourier plane for sampling [35]. In this arrangement, a single pixel detector would be placed at a focal distance of the second lens, while dynamic absorbing masks in the Fourier plane [32, 33] can be used to sample along the circle path. The fastest method, however, would be to use a circular array of THz photoconductive antennas, which have been already demonstrated by several research groups [22-25], thus resulting in a simultaneous interrogation of all the spatial points along a circular path in the k-space.

Finally, according to our methodology, sampling in the second dimension is performed via interpretation of the frequency spectrum of a registered broadband THz pulse. Currently, a slow linear delay line is used for the THz pulse acquisition, thus limiting per pixel pulse acquisition rates to 1 Hz. Slow linear delay lines can be substituted by the much faster rotary delay lines that were recently demonstrated by several research groups [13-21] that can enhance acquisition rate to ~10-100Hz. In fact, mechanical optical delay lines can be replaced altogether by the opto-electronic delay lines integrated into the fs lasers used in the TDS-THz systems, thus resulting in full THz spectral scanning rates of ~100Hz per pixel [12]. In any case, even if the optical delay line can be made infinitely fast, the spectrum acquisition time in the THz-TDS systems is limited by the lock-in averaging time constant of ~10ms, which is a minimal time required per frequency in order to get a reliable intensity reading with a SNR of 20-40dB by amplitude. Using a modest spectral resolution of ~30GHz, a full THz spectrum of up to 3THz can then be acquired in ~1s. Alternatively, tunable THz systems based on frequency difference generation in photomixers can also be used for detection of spectral information [38]. While utilizing technology similar to the photoconductive antennas, and featuring spectral acquisition rates of ~1Hz, such signals offer considerably higher SNRs and signal intensities. Therefore, using linear THz antenna arrays together with fast delay lines (specifically, opto-electronic delay lines) will allow to forgo completely any moving parts in the THz imaging systems based on photoconductive antennas and result in ~1 Hz acquisition rates of the medium resolution ~100x100, very high SNRs THz images either in amplitude contrast or phase contrast modalities. These acquisition rates can be further improved either by sacrificing SNR while using faster optical delay lines, or without sacrificing SNR by using more powerful (albeit considerably more expensive) broadband THz sources such as those based on the two-color fs laser mixing and plasma filamentation (see [39], for example).

## Conclusion

In conclusion, we have demonstrated a novel hybrid Fourier imaging-based reconstruction method that uses both spatial and spectral information to map the spatial frequencies in the Fourier plane. Then, using solid mathematical foundations, we develop theoretically and demonstrate numerically and experimentally application of several algorithms for compression-less reconstruction of high-contrast amplitude images and low-contrast phase images. This work is motivated by the need for a real-time high SNR THz imaging system capable of medium-to-high spatial resolution. We believe that proposed hybrid spatial/spectral image acquisition modality in Fourier plane can enable such systems with already existing technologies.

# Methods

## Experimental setup

The experimental setup is based on a THz-time domain spectroscopy system and depicted in Supplementary Fig. S1. A Ti:Sapphire laser (100 fs, 800 nm, 100 MHz) delivers 300 mW and 10 mW to the THz photoconductive emitter and detector antennae (PCA) respectively. On the emitter side, a linear delay line is placed prior to a high-power interdigitated antenna supplied with 15 V at 5 kHz. In our setup, we use a fiber coupled detector in order to allow convenient scanning of the Fourier plane. In particular, the optical beam is focused into a polarization maintaining optical fiber with the input end rigidly fixed in the fiber coupler on the optical table, while the output end is fixed in a fiber coupler mounted on a 3D micropositioning stage. At the output end of the fiber, the beam is collimated in air before exciting the detection antenna. The optical fiber adds positive group-velocity dispersion (GVD) to the pulse, leading in a non-negligible pulse broadening that reduces the THz bandwidth of the detector. Therefore, before focusing into the fiber, the optical pulse passes through a dispersion pre-compensation system made of two diffraction gratings and a mirror that adds negative GVD to the pulse (see Supplementary Section 1 for more details). The output current of the antenna is amplified with a lock-in amplifier before being recorded by a data acquisition card connected to a computer. Thus, the acquired data is a current proportional to the electric field as a function of time. A temporal Fourier transform is then performed on the pulse to get the amplitude and phase as a function of the THz frequency. Finally, the hybrid inverse transform described in the paper is applied in order to retrieve the field distribution in the object plane.

## A closer look at the image reconstruction algorithm.

The objective of this section is to determine what image is retrieved when using the hybrid inverse transform presented by equation (8). As the integral (7) used in reconstruction integrates information about the object over all frequencies, the reconstructed image $\tilde{S}(\bar{r})$ is different from the original image $S(\bar{r}, \nu)$. In order to understand how $\tilde{S}(\bar{r})$ is related to the original image $S(\bar{r}, \nu)$, we substitute the optical Fourier transform of the image $U(\bar{\rho}, \nu)$ as measured in the Fourier plane of our setup (equation (5)) into the hybrid inverse transform developed in this paper (equation (7)) and get:

$$\tilde{S}(\bar{r}) = \iint d\theta \, \nu d\nu \iint d\phi' \, r' dr' \left(\frac{\rho_0}{cF}\right)^2 \frac{S(\bar{r}', \nu)}{U_{ref}(\nu)} \exp\left[-\frac{j2\pi\nu}{cF} \bar{\rho}_0 \cdot (\bar{r}' - \bar{r})\right] \tag{19}$$

where $r'$ and $\phi'$ are the integration variables that are defined in Supplementary Fig. 2. We can simplify the expressions above by defining $\bar{\varepsilon} = \bar{r}' - \bar{r}$ and $\phi_{\bar{\varepsilon}}$ as shown in Supplementary Fig. S2. Then, equation (19) can be written as:

$$\begin{aligned}\tilde{S}(\bar{r}) &= \iint d\theta \, \nu d\nu \iint d\phi_{\bar{\varepsilon}} \, \varepsilon d\varepsilon \left(\frac{\rho_0}{cF}\right)^2 \frac{S(\bar{r}+\bar{\varepsilon}, \nu)}{U_{ref}(\nu)} \exp\left[-\frac{j2\pi\nu}{cF} \bar{\rho}_0 \cdot \bar{\varepsilon}\right] \\ &= \iint d\theta \, \nu d\nu \iint d\phi_{\bar{\varepsilon}} \, \varepsilon d\varepsilon \left(\frac{\rho_0}{cF}\right)^2 \frac{S(\bar{r}+\bar{\varepsilon}, \nu)}{U_{ref}(\nu)} \exp\left[-\frac{j2\pi\nu}{cF} \rho_0 \varepsilon \right. \\ & \left. \cdot \cos(\theta - \phi_{\bar{\varepsilon}})\right]\end{aligned} \tag{20}$$

The integral above can be further simplified by integrating over $\theta$, thus arriving to the following final expression for the hybrid inverse transform:

$$\tilde{S}(\bar{r}) = 2\pi \int_0^\infty \nu d\nu \int_0^{2\pi} d\phi_{\bar{\varepsilon}} \int_0^\infty \varepsilon d\varepsilon \left[ \left(\frac{\rho_0}{cF}\right)^2 \frac{S(\bar{r}+\bar{\varepsilon},\nu)}{U_{ref}(\nu)} \right] J_0 \left(\frac{2\pi\rho_0}{cF}\nu\varepsilon\right) \quad (21)$$

where we have used the integral representation of the Bessel function from the Supplementary Section 3 (equation S2). In order to continue further, we now need to specify the structure of the original image $S(\bar{r},\nu)$. In the following sections, we consider two particular types of images – those based on amplitude or phase masks.

*Amplitude masks*

First, we consider the case of space-frequency separable amplitude masks that result in the image of the general form $S(\bar{r},\nu) = S(\bar{r})E(\nu)$. We find that when using the reference function $U_{ref}(\nu)$ as defined by equation (11), the hybrid inverse transform (expression (21)) gives the following:

$$\tilde{S}(\bar{r}) = \frac{2\pi}{\iint d\bar{r}\, S(\bar{r})} \left(\frac{\rho_0}{cF}\right)^2 \int_0^{2\pi} d\phi_{\bar{\varepsilon}} \int_0^\infty \varepsilon d\varepsilon\, S(\bar{r}+\bar{\varepsilon}) \int_0^\infty \nu d\nu\, J_0\left(\frac{2\pi\rho_0\varepsilon}{cF}\nu\right) \quad (22)$$

The last integral can be readily evaluated analytically using integral representations of the delta function (equation S4) in terms of the Bessel functions:

$$\int_0^\infty \nu\, d\nu\, J_0\left(\frac{2\pi\rho_0\varepsilon}{cF}\nu\right) = \frac{cF}{2\pi\rho_0\varepsilon} \delta\left(\frac{2\pi\rho_0\varepsilon}{cF}\right) \quad (23)$$

where $\delta(2\pi\rho_0\varepsilon/cF)$ is the Dirac delta function. Then,

$$\tilde{S}(\bar{r}) = \frac{1}{\iint d\bar{r}\, S(\bar{r})} \frac{\rho_0}{cF} \int_0^{2\pi} d\phi_\varepsilon \int_0^\infty d\varepsilon\, S(\bar{r}+\bar{\varepsilon})\, \delta\left(\frac{2\pi\rho_0\varepsilon}{cF}\right) \quad (24)$$

These integrals can then be evaluated analytically by using the properties of the Dirac delta function (equation S5):

$$\tilde{S}(\bar{r}) = \frac{S(\bar{r})}{\iint d\bar{r}\, S(\bar{r})} \quad (25)$$

where the denominator is simply a constant proportional to the area of the image. Therefore, we conclude that in the case of a space-frequency separable amplitude mask, the hybrid inverse transform of the Fourier image results in the original image normalized by the area of the image. If the maximum frequency used in the reconstruction is finite, we can also use equation (22) in order to estimate the resultant image resolution as demonstrated in Supplementary section 4.

*Phase masks*

We now consider the case of phase masks that result in the image of the general form $S(\bar{r},\nu) = S(\bar{r})E(\nu)\exp[j2\pi\nu(\Delta_0 - \mu(\bar{r}))/c]$. As for the choice of the reference function $U_{\text{ref}}(\nu)$ in the inversion algorithm (21), one would be tempted to use the same normalization as in the case of the amplitude masks. However, as we will see in what follows, in order to ensure convergence of all the integrals, we need to modify the reference function. Particularly, in the case of phase masks we define the reference $U_{\text{ref}}(\nu)$ as the Fourier transform of a pulse at the origin of the k-space ($\rho_0 = 0$), measured using a flat reference sample $\mu(\bar{r}) = 0$:

$$U_{\text{ref}}(\nu) = \nu^m \frac{jcF}{\nu} U(0,\nu) = \nu^m E(\nu) \exp[j2\pi\nu\Delta_0/c] \iint d\bar{r}\, S(\bar{r}) \tag{26}$$

where $\nu^m$ is an additional frequency multiplier with an exponent $m$ that must be chosen to ensure the convergence of the inversion algorithm. Using expression (21) for the hybrid inverse transform together with the reference function defined in (26), we find:

$$\tilde{S}(\bar{r}) = \frac{2\pi}{\iint d\bar{r}\, S(\bar{r})} \left(\frac{\rho_0}{cF}\right)^2 \int_0^{2\pi} d\phi_\varepsilon \int_0^\infty \varepsilon d\varepsilon\, S(\bar{r} + \bar{\varepsilon}) K_m(\bar{r}, \bar{\varepsilon}) \tag{27}$$

where we define:

$$K_m(\bar{r}, \bar{\varepsilon}) = \int_0^\infty \nu^{1-m} d\nu \exp[-j2\pi\mu(\bar{r} + \bar{\varepsilon})\nu/c] J_0\left(\frac{2\pi\rho_0\varepsilon}{cF}\nu\right) \tag{28}$$

In the case of amplitude masks considered earlier, we used $m = 0$, however, in the case of phase masks this choice generally results in the divergent integral (27). Indeed, using basic properties of the Bessel function (equation S9) we find that:

$$K_0(\bar{r}, \bar{\varepsilon}) = \left(\frac{c}{2\pi}\right)^2 \left(\frac{F}{\rho_0}\right)^2 \mu_0(\bar{r}) \cdot \begin{cases} \dfrac{-1}{[\mu_0(\bar{r} + \bar{\varepsilon})^2 - \varepsilon^2]^{3/2}} & \text{if } 0 < \varepsilon < \mu_0(\bar{r} + \bar{\varepsilon}) \\ \dfrac{j}{[\varepsilon^2 - \mu_0(\bar{r} + \bar{\varepsilon})^2]^{3/2}} & \text{if } \varepsilon > \mu_0(\bar{r} + \bar{\varepsilon}) \end{cases} \tag{29}$$

where the normalized optical path is defined as $\mu_0(\bar{r} + \bar{\varepsilon}) = \frac{F}{\rho_0}\mu(\bar{r} + \bar{\varepsilon})$. Therefore, the integration over $\varepsilon$ in equation (27) can be divided into the sum of two integrals over different regions of $\varepsilon$ values. From the form of (29) it follows that generally both of these integrals are divergent, thus rendering $\tilde{S}(\bar{r})$ ill defined. Therefore, $m = 0$ is not an acceptable choice for the reference function.

On the other hand, if we select $m = 1$, then, using equation (S10),

$$K_1(\bar{r}, \bar{\varepsilon}) = \frac{1}{2\pi} \frac{cF}{\rho_0} \begin{cases} \dfrac{-j}{\sqrt{\mu_0(\bar{r} + \bar{\varepsilon})^2 - \varepsilon^2}} & \text{if } 0 < \varepsilon < \mu_0(\bar{r} + \bar{\varepsilon}) \\ \dfrac{1}{\sqrt{\varepsilon^2 - \mu_0(\bar{r} + \bar{\varepsilon})^2}} & \text{if } \varepsilon > \mu_0(\bar{r} + \bar{\varepsilon}) \end{cases} \tag{30}$$

where again $\mu_0(\bar{r} + \bar{\varepsilon}) = \frac{F}{\rho_0}\mu(\bar{r} + \bar{\varepsilon})$. Now, we can write equation (27) as the sum of the two integrals:

$$\tilde{S}(\bar{r}) = \frac{2\pi}{\iint d\bar{r}\, S(\bar{r})} \left(\frac{\rho_0}{cF}\right)^2 \int_0^{2\pi} d\phi_\varepsilon \left[S_{\varepsilon<\mu_0(\bar{r}+\bar{\varepsilon})} + S_{\varepsilon>\mu_0(\bar{r}+\bar{\varepsilon})}\right] \tag{31}$$

where

$$S_{\varepsilon<\mu_0(\bar{r}+\bar{\varepsilon})} = -\frac{j}{2\pi}\frac{cF}{\rho_0} \int_0^{\mu_0(\bar{r}+\bar{\varepsilon})} d\varepsilon \frac{\varepsilon \cdot S(\bar{r}+\bar{\varepsilon})}{\sqrt{\mu_0(\bar{r}+\bar{\varepsilon})^2 - \varepsilon^2}} \tag{32}$$

and

$$S_{\varepsilon>\mu_0(\bar{r}+\bar{\varepsilon})} = \frac{1}{2\pi}\frac{cF}{\rho_0} \int_{\mu_0(\bar{r}+\bar{\varepsilon})}^{\infty} d\varepsilon \frac{\varepsilon \cdot S(\bar{r}+\bar{\varepsilon})}{\sqrt{\varepsilon^2 - \mu_0(\bar{r}+\bar{\varepsilon})^2}} \tag{33}$$

We note that the contribution $S_{\varepsilon>\mu_0(\bar{r}+\bar{\varepsilon})}$ is purely real, and it is generally divergent if the function $S(\bar{r}+\bar{\varepsilon})$ (amplitude mask) is not zero at infinity. In the meantime, the contribution $S_{\varepsilon<\mu_0(\bar{r}+\bar{\varepsilon})}$ is purely imaginary and is always finite. Moreover, if the spatial region of the optical path variation $\mu(\bar{r})$ is large enough so its value can be considered constant $\mu_0(\bar{r}+\bar{\varepsilon}) = \mu_0(\bar{r})$ inside a circle of radius $\varepsilon \sim \mu_0(\bar{r})$, and if a similar condition is respected by the amplitude mask $S(\bar{r})$, then we can evaluate equation (33) exactly:

$$\begin{aligned} S_{\varepsilon<\mu_0(\bar{r}+\bar{\varepsilon})} &= -\frac{j}{2\pi}\frac{cF}{\rho_0} \int_0^{\mu_0(\bar{r}+\bar{\varepsilon})} d\varepsilon \frac{\varepsilon \cdot S(\bar{r}+\bar{\varepsilon})}{\sqrt{\mu_0(\bar{r}+\bar{\varepsilon})^2 - \varepsilon^2}} \\ &= -\frac{j}{2\pi}\frac{cF}{\rho_0} S(\bar{r}) \int_0^{\mu_0(\bar{r})} \frac{\varepsilon\, d\varepsilon}{\sqrt{\mu_0(\bar{r})^2 - \varepsilon^2}} = -\frac{j}{2\pi}\frac{cF}{\rho_0} S(\bar{r})\mu_0(\bar{r}) \end{aligned} \tag{34}$$

Therefore, from equation (31), we find that the imaginary part of the hybrid inverse transform $\tilde{S}(\bar{r})$ computed with $m = 1$ is proportional to the spatial variation of the optical path in the image plane:

$$\mathrm{Im}\{\tilde{S}(\bar{r})\} = -\frac{2\pi}{c} \frac{S(\bar{r})\,\mu(\bar{r})}{\iint d\bar{r}\, S(\bar{r})} \tag{35}$$

If the maximum frequency used in the reconstruction is finite, we can also use equation (27) in order to estimate the resultant image resolution as demonstrated in Supplementary section 5.

# Supplementary information

## 1. Additional details of the experimental setup

The detector in the THz-TDS system presented in this work is fiber-coupled (Fig. S1a). This allows to freely move the detector in space without realigning THz setup and record an image via point by point scanning. On the detection line, the beam is focused into a polarization-maintaining optical fiber which has an input end fixed in a fiber coupler mounted on the optical table, while fiber the output end is fixed in the fiber coupler mounted on a 3D micropositioning stage. At the output, the beam is collimated in air before exciting the detection antenna. The optical fiber adds positive group-velocity dispersion (GVD) to the pulse, leading to a non-negligible pulse broadening that reduces the THz bandwidth of the detector. Therefore, before focusing into the fiber, the optical pulse is passed through a dispersion pre-compensation system made of two diffraction gratings and a mirror (Fig. S1b). To counterbalance the fiber's positive GVD, this system adds negative GVD as detailed in [40]:

$$GVD[\text{s}^2] = -\frac{\lambda^3 d}{\pi c^2 \Lambda^2}\left[1-\left(\frac{\lambda}{\Lambda}-\sin\theta\right)^2\right]^{3/2} \quad \text{(S1)}$$

where $\lambda$ is the central wavelength, $c$ is the speed of light, $\Lambda$ is the groove distance of the gratings, $d$ is the distance between the gratings and $\theta$ is the incident angle defined in Fig. S1b. By adjusting the geometrical parameters $d$ and $\theta$, it is possible to counterbalance the positive GVD of the fiber, and therefore minimize the pulse duration at the output of the fiber. In Fig. S1c, we present several autocorrelation traces of the pulse for different values of $d$, recorded using an autocorrelator kit from Newport. As it can be seen in Fig. S1d, the pulse duration is minimized to 213 fs when using $d = 29.5$ mm between the gratings. We note that the compensation system only corrects for positive GVD. In the meantime, negative GVD due to the third order dispersion or non-linear effects still occur in the fiber and are not compensated by our system. Therefore, in our system, we still have a pulse broadening. Also, in principle, one could also use dispersion compensation fiber at 1550 nm since THz emitters and detectors are also commercially available at that wavelength.

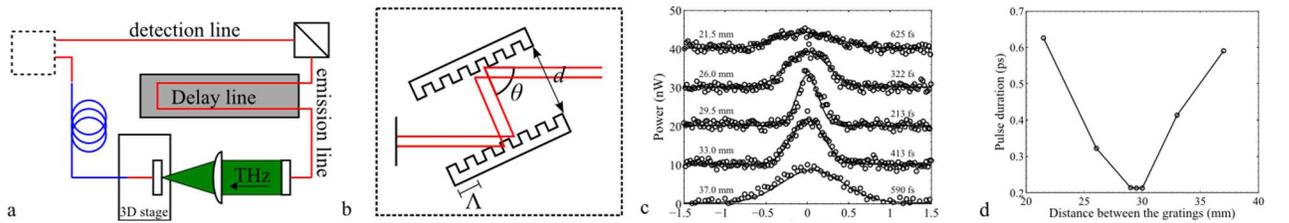

**Figure S1. Experimental setup.** (a) Fiber-coupled THz-TDS. The red lines refer to the free-space beam, while the blue line is the optical fiber. (b) Dispersion compensation system with two diffraction gratings. (c) Autocorrelation traces for different distances between the gratings. The curves are fitted with a Gaussian function. (d) Pulse duration (full width at half maximum) as a function of the distance between the gratings.

## 2. Definition of various integration variables

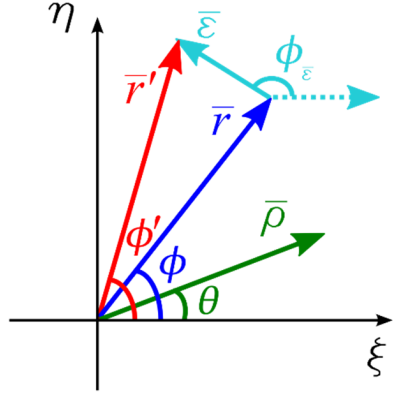

**Figure S2. Definition of various integration variables used throughout the paper.**

## 3. Mathematical identities

In this section, we list various mathematical identities used throughout the paper.

*Integral representation of the Bessel function of a real argument and some of its basic properties:*

$$J_0(x) = \frac{1}{2\pi}\int_0^{2\pi} e^{\pm ix\cos\theta}d\theta \tag{S2}$$

$$J_1(ax) = -\frac{1}{a}J_0'(ax) \tag{S3}$$

*Dirac delta function, properties and integral representations:*

$$\delta(x) = \int_{-\infty}^{\infty} e^{j2\pi x\xi}d\xi \tag{S4}$$

$$\delta(ax) = \frac{1}{|a|}\delta(x) \tag{S5}$$

$$\int_0^{\infty} xJ_0(ax)dx = \frac{1}{a}\delta(a) \tag{S6}$$

$$\int_{-\infty}^{\infty} f(x)\delta(x) = f(0) \tag{S7}$$

*Some integrals involving Bessel functions:*

$$\int_0^x uJ_0(au)du = \frac{x}{a}J_1(ax) \tag{S8}$$

$$\int_0^{\infty} xe^{-jax}J_0(bx)dx = \begin{cases} \dfrac{ja}{(b^2-a^2)^{3/2}}, & |b|>|a| \\ \dfrac{-a}{(a^2-b^2)^{3/2}}, & |b|<|a| \end{cases} \tag{S9}$$

$$\int_0^{\infty} e^{-jax}J_0(bx)dx = \begin{cases} \dfrac{1}{\sqrt{b^2-a^2}}, & |b|>|a| \\ -\dfrac{j\cdot\mathrm{sgn}(a)}{\sqrt{a^2-b^2}}, & |b|<|a| \end{cases} \tag{S10}$$

## 4. Resolution in the case of amplitude masks

Image resolution in the case of amplitude masks, can be readily deduced from the hybrid inverse transform equation (22):

$$\tilde{S}(\bar{r}) = \frac{2\pi}{\iint d\bar{r}\, S(\bar{r})} \left(\frac{\rho_0}{cF}\right)^2 \int_0^{2\pi} d\phi_{\bar{\varepsilon}} \int_0^{\infty} \varepsilon d\varepsilon\, S(\bar{r}+\bar{\varepsilon}) \int_0^{\infty} v dv\, J_0\left(\frac{2\pi \rho_0 \varepsilon}{cF} v\right) \tag{S11}$$

We note that, as the probing pulse has a finite duration (bandwidth limited pulse), the maximal frequency in the integral presented above is limited to $v_{max}$. This also sets the minimal resolution achievable by the hybrid inverse transform, which as we show below is proportional to the smallest probed wavelength ($\lambda_{min} = c/v_{max}$). To demonstrate this point mathematically, we consider the last integral in equation (S11), however, instead of integration to infinity, we now integrate up to $v_{max}$. In this case, using the Bessel identity (equation (S6)), the last integral of equation (S11) becomes:

$$\int_0^{v_{max}} v\, dv\, J_0\left(\frac{2\pi\rho_0\varepsilon}{cF} v\right) = \frac{cF v_{max}}{2\pi\rho_0\varepsilon} J_1\left(\frac{2\pi\rho_0\varepsilon}{cF} v_{max}\right) \tag{S12}$$

Then, by using the property (S7), equation (S11) transforms into:

$$\begin{aligned}
\tilde{S}(\bar{r}) &= \frac{v_{max}}{\iint d\bar{r}\, S(\bar{r})} \frac{\rho_0}{cF} \int_0^{2\pi} d\phi_\varepsilon \int_0^{\infty} d\varepsilon\, S(\bar{r}+\bar{\varepsilon}) J_1\left(\frac{2\pi\rho_0 v_{max}}{cF}\varepsilon\right) \\
&= -\frac{1}{2\pi \iint d\bar{r}\, S(\bar{r})} \int_0^{2\pi} d\phi_\varepsilon \int_0^{\infty} d\varepsilon\, S(\bar{r}+\bar{\varepsilon}) J_0'\left(\frac{2\pi\rho_0 v_{max}}{cF}\varepsilon\right) \\
&= -\frac{1}{2\pi \iint d\bar{r}\, S(\bar{r})} \int_0^{2\pi} d\phi_\varepsilon \int_0^{\infty} S(\bar{r}+\bar{\varepsilon})\, dJ_0\left(\frac{2\pi\rho_0 v_{max}}{cF}\varepsilon\right) \\
&= \frac{1}{2\pi \iint d\bar{r}\, S(\bar{r})} \int_0^{2\pi} d\phi_\varepsilon \left[S(\bar{r}) + \int_0^{\infty} d\varepsilon\, \frac{\partial S(\bar{r}+\bar{\varepsilon})}{\partial \varepsilon} J_0\left(\frac{2\pi\rho_0 v_{max}}{cF}\varepsilon\right)\right] \\
&= \frac{S(\bar{r})}{\iint d\bar{r}\, S(\bar{r})} \\
&\quad + \frac{1}{2\pi \iint d\bar{r}\, S(\bar{r})} \int_0^{2\pi} d\phi_\varepsilon \int_0^{\infty} d\varepsilon\, \frac{\partial S(\bar{r}+\bar{\varepsilon})}{\partial \varepsilon} J_0\left(\frac{2\pi\rho_0 v_{max}}{cF}\varepsilon\right)
\end{aligned} \tag{S13}$$

From equation (S13), we note that the leading expression for the hybrid inverse transform in the case of a bandwidth limited pulse is still proportional to the value of the original image, with a correction term featuring an integration of the amplitude mask spatial derivative over a fast-varying sign-changing Bessel function of a large argument. The main contribution to this integral comes from averaging the amplitude mask derivative inside of a circle of radius $\varepsilon \sim \lambda_{min} \frac{F}{\rho_0}$. From this we conclude that the fundamental limit of the spatial resolution in the case of the amplitude mask is set by the finite bandwidth of the probing pulse and is $\sim \lambda_{min} \frac{F}{\rho_0}$, where $\lambda_{min}$ is the smallest wavelength probed by the pulse. This result correlates well with the Nyquist theorem for the Fourier transform discussed in the paper.

To confirm this result we perform numerical simulations using as an object a binary mask featuring a pattern of lines of increasing size of ones (full transmittance) and zeros (full absorbance) (Fig. S3a). In Fig. S3a, the smallest line of "ones" has a width of 400 µm followed by an adjacent line of "zeros" of the same size. The pairs of lines are then repeated with increasing size in increments of 400 µm (i.e. 400 µm, 800 µm, 1200 µm etc.). We then compute the direct transform given by equation (1) for all the frequencies specified by the THz pulse bandwidth (0.1 - 2 THz with a step of 1 GHz). Then, using the hybrid inverse transform (equation (12)) we reconstruct the original image. In Fig. S3b to S3f, we present several reconstructed images for different values of the maximal THz frequency used in the hybrid inverse transform. In Fig. S4, we perform the same numerical simulations for a geometrically similar image that feature thinner lines (a minimal width of 100 µm and a step of 200 µm in the width increment between the pairs of the adjacent lines). Just by looking at the images, as expected, we find qualitatively that the resolution improves when the THz bandwidth increases.

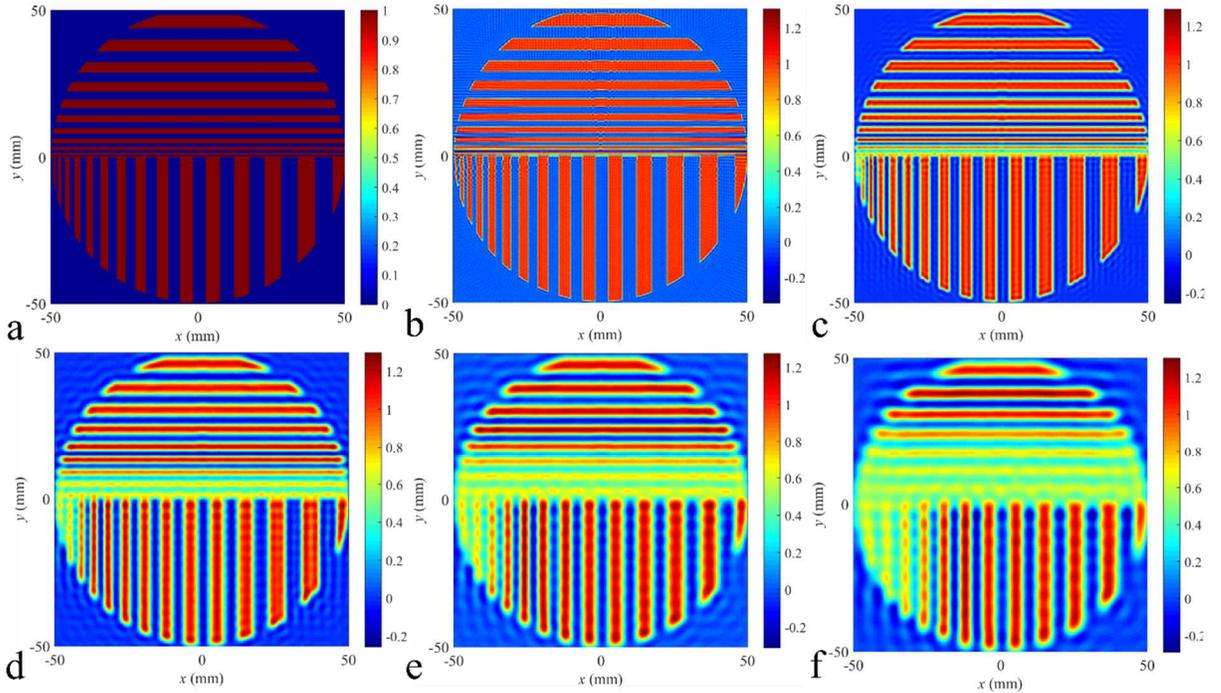

**Figure S3. Numerical reconstruction of a binary mask of 400 µm minimal line width.** (a) Target image featuring pairs of "ones" and "zeros" lines of increasing widths. The smallest line width is 400 µm, the increment in the line width is also 400 µm. (b) Image reconstruction using hybrid inverse transform (equation (12)) with $\lambda_{min} = 150$ µm ($\nu_{max} = 2$ THz), (c) $\lambda_{min} = 350$ µm ($\nu_{max} = 0.86$ THz), (d) $\lambda_{min} = 600$ µm ($\nu_{max} = 0.5$ THz), (e) $\lambda_{min} = 800$ µm ($\nu_{max} = 0.375$ THz) and (f) $\lambda_{min} = 1000$ µm ($\nu_{max} = 0.3$ THz).

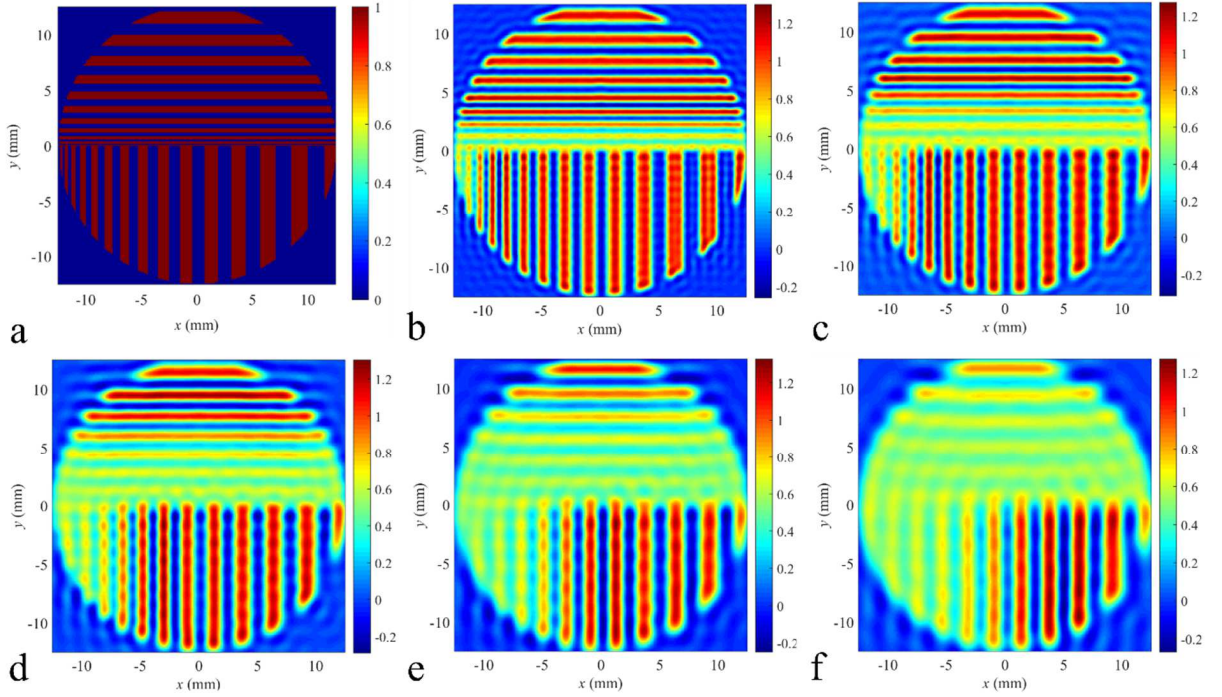

**Figure S4. Numerical reconstruction of a binary mask of 100 μm minimal line width.** (a) Target image featuring pairs of "ones" and "zeros" lines of increasing widths. The smallest line width is 100 μm, the increment in the line width is also 100 μm. (b) Image reconstruction using hybrid inverse transform (equation (12)) with $\lambda_{min}$ = 150 μm ($\nu_{max}$ = 2 THz), (c) $\lambda_{min}$ = 200 μm ($\nu_{max}$ = 1.5 THz), (d) $\lambda_{min}$ = 250 μm ($\nu_{max}$ = 1.2 THz), (e) $\lambda_{min}$ = 300 μm ($\nu_{max}$ = 1 THz) and (f) $\lambda_{min}$ = 350 μm ($\nu_{max}$ = 0.86 THz).

From these numerical results, we can estimate dependence of the spatial resolution of images obtained using the hybrid inverse transform (12) as a function of the maximal THz frequency used in the reconstruction. To do so, for each line of "zeros" and "ones", we calculate the average reconstructed intensities $I_1$ and $I_0$ at the positions where the corresponding lines should be. We then plot these averaged values for each line. If the line of "ones" is resolved one should thus find an averaged value close to one at the line location. Similarly, if the line of "zeros" is resolved one should find an averaged value close to zero at the corresponding line location. The adjacent "zero" and "one" lines are considered not resolved if the corresponding averaged values for these lines become comparable, that is $(I_1 - I_0)/(I_1 + I_0) < 0.5$. To illustrate this, in Fig. S5a, we present the averaged intensities at the locations of "zeros" and "ones" in the case of $\lambda_{min}$ = 600 μm ($\nu_{max}$ = 0.5 THz). The lines from 400 μm to 1600 μm are not resolvable since "zeros" and "ones" have comparable averaged values. However, starting with a 2000 μm thick line, the lines can be readily resolved according to the condition presented above. From this we conclude that in the case of $\lambda_{min}$ = 600 μm, the minimal achievable resolution is ~2000 μm.

We then perform the same analysis for other values of $\lambda_{min}$ (various THz pulse bandwidths) and present the results in Fig. S5b. The dotted line corresponds to the resolution set by the Nyquist theorem : $\delta x = 0.5\,\lambda_{min} F/\rho_0$, while in circles we present resolutions found using the approach described above. From this, we confirm that the resolution of the binary amplitude mask indeed follows closely prediction of the Nyquist theorem. This is further confirmed experimentally in Fig. S5c. There, we use three different binary masks in the form of metal plates with cutouts in the shape of lines. Each amplitude mask features three cutouts in the form of lines of 15 mm in length and widths of 2800 µm, 2000 µm and 1200 µm that are separated by the same width from each other. Then, using our Fourier optics setup and the hybrid inverse algorihm we reconstruct the images of the three lines using different values of the maximal frequency $\nu_{max}$ in the integral (12) (different THz bandwidths). The inserts i, ii, iii in the top row of Fig. S5c show images of the three different amplitude masks that were reconstructed using $\lambda_{min}$ (850 µm, 600 µm, 300 µm respectively). According to the Nyquist theorem, these wavelengths are small enough to correctly resolve the lines in thecorresponding amplitude masks, which is indeed what is observed experimentally. At the same time, in the bottom row of Fig. S5c (inserts iv, v, vi), we show images of the same three amplitude masks that were reconstructed using $\lambda_{min}$ of 1100 µm, 750 µm, and 500 µm respectively. According to the Nyquist theorem, these values of $\lambda_{min}$ are insufficient to resolve the lines in the coresponding amplitude masks, which is indeed what is confirmed experimentally. From this we conclude that, indeed, our analysis of the image resolution in the case of amplitude masks agrees well with the classic Nyquist theorem applied to the classical Fourier optics systems.

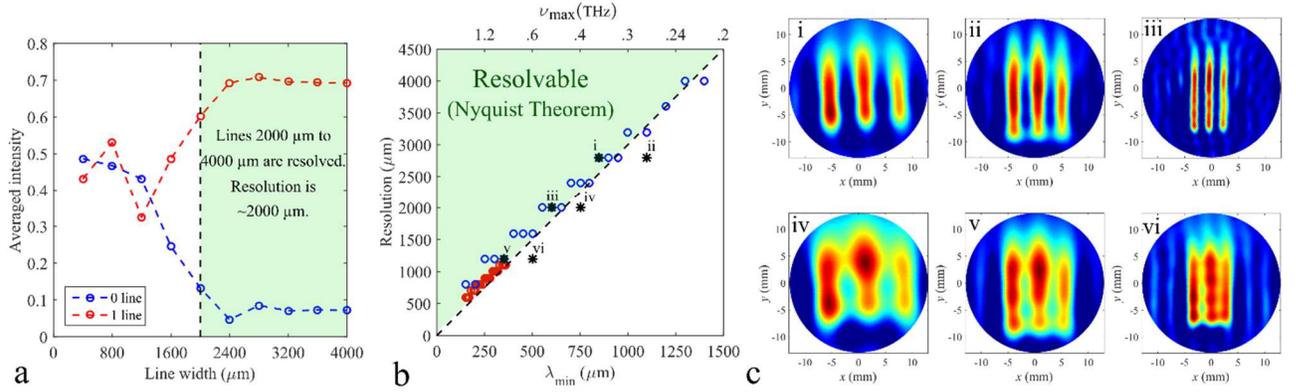

**Figure S5. Image resolution, case of the amplitude masks.** (a) Definition of the resolution using averaged values at the location of "ones" and "zeros". (b) Resolution of amplitude masks as a function of the THz bandwidth. The green region corresponds to the resolvable region set by the Nyquist theorem. The blue and red circles correspond to Fig. S3 and S4 respectively. (c) Experimental measurements using three different amplitude masks featuring three cutout lines. Reconstructed images of the amplitude masks as a function of $\lambda_{min}$ ($\nu_{max}$). i. Lines of 2800 µm with $\lambda_{min} = 850$ µm ($\nu_{max} = 0.35$ THz) and iv. $\lambda_{min} = 1100$ µm ($\nu_{max} = 0.27$ THz); ii. lines of 2000 µm with $\lambda_{min} = 600$ µm ($\nu_{max} = 0.5$ THz) and v. $\lambda_{min} = 750$ µm ($\nu_{max} = 0.4$ THz); iii. lines of 1200 µm with $\lambda_{min} = 300$ µm ($\nu_{max} = 1$ THz) and vi. $\lambda_{min} = 500$ µm ($\nu_{max} = 0.6$ THz).

## 5. Resolution in the case of phase masks

In the case of phase masks, we have demonstrated earlier that the imaginary part of the hybrid inverse transform is, in some limit, simply proportional to the local optical path incurred by the light due to passage through the phase mask :

$$\text{Im}\{\tilde{S}(\bar{r})\} = -\frac{1}{\iint d\bar{r}\, S(\bar{r})} \frac{\rho_0}{cF} \int_0^{2\pi} d\phi_\varepsilon \int_0^{\mu_0(\bar{r}+\bar{\varepsilon})} d\varepsilon \frac{\varepsilon \cdot S(\bar{r}+\bar{\varepsilon})}{\sqrt{\mu_0(\bar{r}+\bar{\varepsilon})^2 - \varepsilon^2}} \approx -\frac{2\pi}{c} \frac{S(\bar{r})\, \mu(\bar{r})}{\iint d\bar{r}\, S(\bar{r})} \quad (S14)$$

This result is strictly valid in the case of piecewise constant phase masks if the value of the normalized optical path $\mu_0(\bar{r})$ and the value of the amplitude mask $S(\bar{r})$ are constant within the circle of radius $\varepsilon = \mu_0(\bar{r}+\bar{\varepsilon})$.This result, however, breaks in the vicinity of the boundaries between different optical path regions of the phase mask. We therefore conclude that in the case of a hybrid inverse transform applied to the phase masks, there is a fundamental limit to the resolution of such a transform which is set by the local value of the normalized optical path length $\mu_0(\bar{r}) = \mu(\bar{r})\frac{F}{\rho_0} = h(\bar{r})(n_m - n_a)$. Therefore, in the case of phase masks, a correction term needs to be added to the resolution given by the Nyquist theorem:

$$\delta x = [0.5\,\lambda_{\min} + h(n_m - n_a)] \cdot \frac{F}{\rho_0} \tag{S15}$$

To confirm this prediction, we perform numerical simulations using as target images phase masks featuring geometrical patterns of lines identical to those of amplitude masks presented in Figs. S3 and S4. In the case of phase masks, however, we consider as lines shallow scratches of depth $h$ engraved onto a plastic plate of refractive index $n_m = 1.62$ (Fig. S6a). For the reconstruction of the engraving pattern we use the hybrid inverse transform (17) corresponding to the case of the phase masks. In Fig. S6b to S6f, we present the reconstructed optical paths for different values of the line depth $h$ and the same THz bandwidth ($\lambda_{\min}$ fixed to 150 µm). As expected from (S15), the resolution improves as the scratch depth decreases.

With a definition of the resolution similar to the binary amplitude mask ("ones" are the lines corresponding to engravings, "zeros" are the lines where there is no engraving), we numerically find the resolution for different THz bandwidths and optical thicknesses. In Fig. S7a, we plot the resolution as a function of the THz bandwidth for different values of the engraving depth. The dotted line in the figure has the equation $\delta x = 0.5\,\lambda_{\min}\,F/\rho_0$ and corresponds to the fundamental limit set by the Nyquist theorem. From the figure, we confirm that the minimal resolution is indeed set by the Nyquist theorem. At the same time, the resolution can be considerably worse than that predicted by the Nyquist theorem due to the contribution proportional to the optical path length of the engraving, which can be significant for deep engravings. Thus, in Fig. S7b, we plot this limit as a function of the engraving depths. There, the dotted line follows the equation $\delta x = h(n_m - n_a)\,F/\rho_0$ and it defines the fundamental limit to the resolution in case of the phase masks in the limit of an infinite pulse bandwidth.

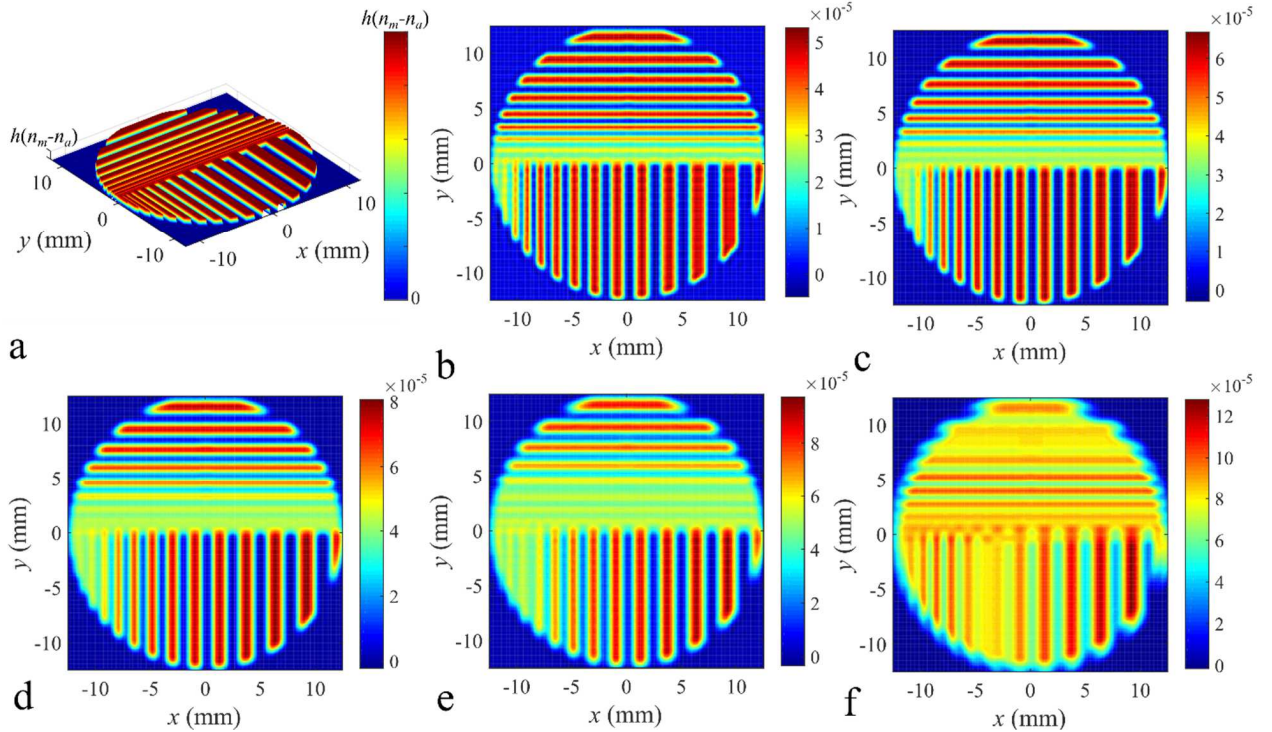

**Figure S6. Numerical reconstruction of a phase mask of 100 µm minimal line width.** (a) Target image featuring pairs of "ones" and "zeros" lines of increasing widths. The smallest line width is 100 µm, the increment in the line width is also 100 µm. The lines of "ones" are the scratches of depth $h$, while the lines of "zeros" correspond to the unperturbed substrate. Reconstructed optical path variation across the substrate for a fixed bandwidth $\lambda_{min} = 150$ µm ($\nu_{max} = 2$ THz) and variable scratch depths (b) $h = 75$ µm, (c) $h = 100$ µm, (d) $h = 125$ µm, (e) $h = 150$ µm and (f) $h = 250$ µm.

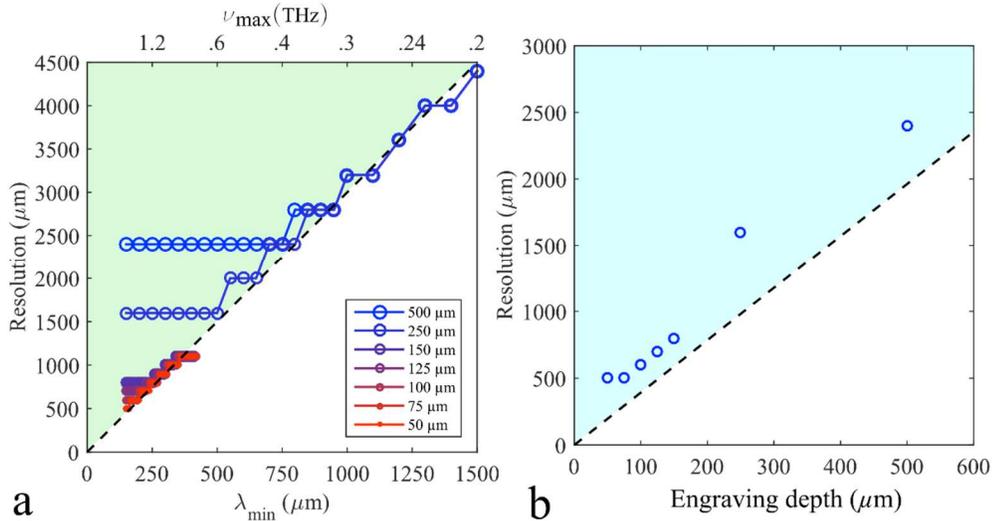

**Figure S7. Image resolution, case of the phase masks.** (a) Resolution as a function of the THz bandwidth for different values of the engraving depth $h$ (shown in the legend). The green region corresponds to the resolvable region set by the Nyquist theorem $\delta x > 0.5 \, \lambda_{min} F/\rho_0$. (b) Resolution as a function of the engraving depth $h$ for a fixed THz bandwidth ($\lambda_{min} = 150$ µm, $\nu_{max} = 2$ THz). The blue region corresponds to the resolvable region as defined by the equation $\delta x > h(n_m - n_a) F/\rho_0$.

# References

1. B. B. Hu, and M. C. Nuss, "Imaging with terahertz waves," *Optics Letters*, vol. 20, no. 16, pp. 1716-1718 (1995).
2. P. U. Jepsen, D. G. Cooke, and M. Koch, "Terahertz spectroscopy and imaging – Modern techniques and applications," *Laser and Photonics Review*, vol. 5, no. 1, pp. 124-166 (2011).
3. J. F. Federici, B. Schulkin, F. Huang, D. Gary, R. Barat, F. Oliveira, and D. Zimdars, "THz imaging and sensing for security applications – explosives, weapons and drugs," *Semiconductor Science and Technolgy*, vol. 20, no. 7, pp. S266-S280 (2005).
4. E. Pickwell, and V. P. Wallave, "Biomedical applications of terahertz technology," *Journal of Physics D: Applied Physics*, vol. 39, no. 17, pp. R301-R310 (2006).
5. A. J. Fitzgerald, B. E. Cole, and P. F. Taday, "Nondestructive analysis of tablet coating thicknesses using terahertz pulsed imaging," *Journal of Pharmaceutical Sciences*, vol. 94, no. 1, pp. 177-183 (2004).
6. A. A. Gowen, C. O'Sullivan, and C. P. O'Donnell, "Terahertz time domain spectroscopy and imaging: Emerging techniques for food process monitoring and quality control," *Trends in Food Science and Technology*, vol. 25, no. 1, pp. 40-46 (2012).
7. E. Abraham, A. Younus, J. C. Delagnes, and P. Mounaix, "Non-invasive investigation of art paintings by terahertz imaging," *Applied Physics A*, vol. 100, no. (3), 585-590 (2010).
8. Y.-S. Lee, *Principles of Terahertz Science and Technology*. Springer, 2009, ch. Generation and Detection of Broadband Terahertz Pulses.
9. Q. Wu, T. D. Hewitt, and X.-C. Zhang. "Two-dimensional electro-optic imaging of THz beams," *Applied Physics Letters*, vol. 69, no. 8, pp. 1026-1028, 1996.
10. S.-G. Park, M. R. Melloch, A. M. Weiner, "Analysis of terahertz waveforms measured by photoconductive and electrooptic sampling," *IEEE Journal of Quantum Electronics*, vol. 35, no. 5, pp. 810-819, 1999.
11. Z. Jiang, and X.C. Zhang, *Sensing with Terahertz Radiation*, Springer, 2013, ch. Free-Space Electro-Optic Techniques.
12. A. Bartels, R. Cerna, C. Kistner, A. Thoma, F. Hudert, C. Janke, and T. Dekorsy, "Ultrafast time-domain spectroscopy based on high-speed asynchronous optical sampling," *Review of Scientific Instruments*, vol. 78, no. 3, p. 035107, 2007.
13. J. Ballif, R. Gianotti, P. Chavanne, R. Wälti, and R. P. Salathé, "Rapid and scalable scans at 21 m/s in optical low-coherence reflectometry," *Optics Letters.*, vol. 22, no. 11, pp. 757–759, 1997.
14. J. Szydlo, N. Delachenal, R. Gianotti, R. Walti, H. Bleuler, and P. R. Salathe, "Air-turbine driven optical low coherence reflectometry at 28.6-kHz scan repetition rate," *Optics Communication*, vol. 154, no. 1–3, pp. 1–4, 1998.
15. G G. Lamouche, M. Dufour, B. Gauthier, V. Bartulovic, M. Hewko, and J.-P. Monchalin, "Optical delay line using rotating rhombic prisms," *Proceedings of the SPIE* vol. 6429, 2007
16. T. Probst, A. Rehn, S. F. Busch, S. Chatterjee, M. Koch, and M. Scheller, "Cost-efficient delay generator for fast terahertz imaging," *Optics Letters*, vol. 39, no. 16, pp. 4863–4866, 2014.
17. D. C. Edelstein, R. B. Romney, and M. Scheuermann, "Rapid programmable 300 ps optical delay scanner and signal averaging system for ultrafast measurements," *Review of Scientific Instruments*, vol. 62, no. 3, pp.579–583, 1991.
18. C.-L. Wang and , C.-L. Pan, National Science Council, Taiwan, R.O.C," Scanning optical delay device having a helicoid reflecting mirror," U.S. Patent 5907423 A, 1999.
19. G.-J. Kim, S.-G. Jeon, J.-I. Kim, and Y.-S. Jin, "High speed scanning of terahertz pulse by a rotary optical delay line," *Review of Scientific Instruments*, vol. 79, no. 10, p. 106102, 2008.
20. M. Skorobogatiy, "Linear rotary optical delay lines," *Optics Express*, vol. 22, no. 10, pp. 11812–11833, 2014.
21. H. Guerboukha, A. Markov, H. Qu, and M. Skorobogatiy, "Time resolved dynamic measurements at THz frequencies using a rotary optical delay line," *IEEE Transactions on Terahertz Science and Technology*, vol. 5, no. 4, 2015.
22. B. Pradarutti, R. Müller, W. Freese, G. Mattahaüs, S. Riehemann, G. Notni, S. Nolte, and A. Tünnermann, "Terahertz line detection by a microlens array coupled photoconductive antenna array," *Optics Express*, vol. 16, no. 22, pp. 18443-18450, 2008.

23. A. Brahm, A. Wilms, R. J. B. Dietz, T. Göbel, M. Schell, G. Notni, an A. Tünnermann, "Multichannel terahertz time-domain spectroscopy system at 1030 nm excitation wavelength," *Optics Express*, vol. 22, no. 11, pp. 12982-12993, 2014.
24. K. Nallappan, J. Li, H. Guerboukha, A. Markov, B. Petrov, D. Morris, and M. Skorobogatiy, "A dynamically reconfigurable terahertz array antenna for near-field imaging applications," *ArXiv*:1705.10624, (2017).
25. K. Nallappan, J. Li, H. Guerboukha, A. Markov, B. Petrov, D. Morris, and M. Skorobogatiy, "A dynamically reconfigurable array antenna for 2D-imaging applications", *Photonics North*, paper 36.30, Canada (2017).
26. Z. Jiang, and X.-C. Zhang, "Terahertz imaging via electrooptic effect," *IEEE Transactions on Microwave Theory and Techniques*, vol. 47, no. 12, 1999.
27. Z. Jiang, X. G. Xu, and X.-C. Zhang, "Improvement of terahertz imaging with a dynamic subtraction technique," *Applied Optics*, vol. 39, no. 17, pp. 2982-2987 (2000).
28. D. L. Donoho, "Compressed sensing," *IEEE Transactions on information theory*, vol. 52, no. 4, pp.1286-1306, 2006.
29. W. L. Chan, M. L. Moravec, R. G. Baraniuk and D. M. Mittleman, "Terahertz imaging with compressed sensing and phase retrieval," *Optics Letters*, vol. 33, no. 9, 974-976 (2008).
30. W. L. Chan, K. Charan, D. Takhar, K. F. Kelly, R. G. Baraniuk, and D. M. Mittleman, "A single-pixel terahertz imaging system based on compressed sensing," *Applied Physics Letters*, vol. 93, no. 12 121105 (2008).
31. Y. C. Chen, L. Gan, M. Stringer, A. Burnett, K. Tych, H. Shen, J. E. Cunningham, E. P. J. Parrot, J. A. Zeitler, L. F. Gladden, E. H. Linfield, and A. G. Davies, "Terahertz pulsed spectroscopic imaging using optimized binary masks," *Applied Physics Letters*, vol. 95, no. 23, 2010.
32. H. Shen, L. Gan, N. Newman, Y. Donc, C. Li, Y. Huang, and Y. C. Shen, "Spinning disk for compressive imaging," *Optics Letters*, vol. 37, no. 1, pp.46-48, 2012.
33. C. M. Watts, D. Shrekenhamer, J. Montoya, G. Lipworth, J. Hunt, T. Sleasman, S. Krishna, D. R. Smith, and W. J. Padilla, "Terahertz compressive imaging with metamaterial spatial light modulators," *Nature Photonics*, vol. 8, pp. 605-609, 2014.
34. D. Shrekenhamer, C. M. Watts, and W. J. Padilla, "Terahertz single-pixel imaging with an optically controlled dynamic spatial light modulator," *Optics Express*, vol. 21, no. 10, pp. 12507-12518, 2013.
35. K. Lee, K. H. Jin, J. C. Ye, and J. Ahn, "Coherent optical computing for T-ray imaging," *Optics Letters*, vol. 35, no. 4, 2010.
36. J. W. Goodman, Introduction to Fourier Optics, W. H. Freeman & Co Ltd., 2005.
37. J. Li, K. Nallappan, H. Guerboukha, and M. Skorobogatiy, "3D printed hollow core terahertz Bragg waveguides with defect layers for surface sensing applications," *Optics Express*, vol. 25, pp. 4126-4144, 2017.
38. S. Preu, G. H. Döhler, S. Malzer, L. J. Wand, and A. C. Gossard, "Tunable, continuous-wave terahertz photomixer sources and applications," *Journal of Applied Physics*, vol. 109, no. 6 (2011).
39. F. Blanchard, G. Sharma, X. Ropagnol, L. Razzari, R. Morandotti, and T. Ozaki, "Improved terahertz two-color plasma sources pumped by high intensity laser beam," *Optics Express*, vol. 17, no. 8, pp. 6044–6052 (2009).
40. F. Ellrich, T. Weinland, D. Molter, J. Jonuscheit, and R. Beigang, "Compact fiber-coupled terahertz spectroscopy system pumped at 800 nm wavelength," *Review of Scientific Instruments*, vol. 82, no. 5, 2011.